\documentclass{svjour2}               

\smartqed  

\usepackage{latexsym}
\usepackage{bm,bbm}
\usepackage{amsmath,amsfonts,amscd,amssymb}
\usepackage{amstext}
\usepackage{graphicx}
\usepackage{verbatim}
\usepackage{graphicx,color}

 \journalname{
}
\newcommand{\be}{\begin{equation}}
\newcommand{\ee}{\end{equation}}
\newcommand{\barr}{\begin{eqnarray}}
\newcommand{\earr}{\end{eqnarray}}
\newcommand{\EE}{\mathcal{E}}
\newcommand{\A}{\mathcal{A}}
\newcommand{\ZZ}{\mathcal{Z}}
\newcommand{\SSS}{\mathcal{S}}

\newcommand{\R}{\mathbb{R}}
\newcommand{\C}{\mathbb{C}}

\newcommand{\de}{\mathrm{d}}

\renewcommand{\rho}{\varrho}

\newcommand{\PP}{\mathcal{P}}
{\left\lbrace\begin{array}{@{}l@{}}}%
{\end{array}\right.}

\begin{document}

\title{Large deviations of radial statistics in the two-dimensional one-component plasma}

\author{\mbox{Fabio Deelan Cunden \and  Francesco Mezzadri \and  Pierpaolo Vivo}}


\institute{Fabio Deelan Cunden \at
              School of Mathematics, University of Bristol, University Walk, Bristol BS8 1TW, United Kingdom\\
              \email{fabiodeelan.cunden@bristol.ac.uk}          
           \and
           Francesco Mezzadri \at
              School of Mathematics, University of Bristol, University Walk, Bristol BS8 1TW, United Kingdom\\
\email{F.Mezzadri@bristol.ac.uk}   
\and Pierpaolo Vivo \at
King's College London, Department of Mathematics, Strand, London WC2R 2LS, United Kingdom\\
\email{pierpaolo.vivo@kcl.ac.uk}
}

\date{
}

\maketitle

\begin{abstract}
The two-dimensional one-component plasma is an ubiquitous model for several vortex systems. For special values of the coupling constant $\beta q^2$ (where $q$ is the particles charge and $\beta$ the inverse temperature), the model also corresponds to the eigenvalues distribution of normal matrix models. Several features of the system are discussed in the limit of large number $N$ of particles for generic values of the coupling constant. We show that the statistics of a class of radial observables produces a
rich phase diagram, and their asymptotic behaviour in terms of large deviation functions is calculated explicitly, including next-to-leading terms up to order $1/N$. We demonstrate a split-off phenomenon associated to atypical fluctuations of the edge density profile. We also show explicitly that a failure of the fluid phase assumption of the plasma can break a genuine $1/N$-expansion of the free energy. Our findings are corroborated by numerical comparisons with exact finite-$N$ formulae valid for $\beta q^2=2$.
\end{abstract}

\section{Introduction and main results}\label{sec:intro}
In recent years there has been a considerable interest in the study of systems with logarithmic interactions. The simplest example of models in this class is the two-dimensional one-component plasma (2D-OCP). This system is also known in literature as `jellium', 2D Dyson's gas or 2D Coulomb gas~\cite{Alastuey81,Cornu88,Dyson62,Forrester98,Jancovici81,Johannesen83,Sari76a,Weigmann06}. The 2D-OCP consists of $N$ identical classical
point-like particles, each carrying a charge $q$ (one species of particle) on a two-dimensional domain. The Coulomb interaction between any two particles at distance $\vec{r}$ from each other is $-q^2v(\vec{r})$, where $v(\vec{r})$ obeys the Poisson equation. In the planar case $v(\vec{r})=\log(|\vec{r}|/L)$, where $L$ is a length scale that fixes the zero of the potential.
To ensure charge neutrality, the particles are embedded in a fixed neutralizing background with opposite charge $-qN$. 
The canonical distribution of the 2D-OCP at inverse temperature $\beta$ is
\barr
&&\mathbb{P}_{\beta,N}\left(\vec{r}_1,\dots,\vec{r}_N\right)=\displaystyle\frac{1}{\mathcal{Z}_{N,\beta}}e^{-\beta H\left(\vec{r}_1,\dots,\vec{r}_N\right)}\ ,\label{eq:jpdf}\\
&&H\left(\vec{r}_1,\dots,\vec{r}_N\right)=-\displaystyle\frac{q^2}{2}\sum_{i\neq j}\log\left(\frac{|\vec{r}_i-\vec{r}_j|}{L}\right)+q^2N\sum_kV\left(\frac{\vec{r}_k}{L}\right)\ . \label{eq:hamiltonian}
\earr
In \eqref{eq:hamiltonian}, $\vec{r}_i=(x_i,y_i)\in\R^2$ denotes the position of the $i$-th particle of the 2D-OCP ($i=1,\dots,N$) and $|\cdot|$ denotes the Euclidean distance. The first term in $H\left(\vec{r}_1,\dots,\vec{r}_N\right)$ is the particle-particle contribution to the energy, while the second term is the particle-background contribution (the 2D-OCP experiences the electrostatic potential $V$ generated by the fixed background). The coupling constant $\beta q^2$ is often referred to as \emph{plasma parameter}. 

This statistical mechanics fluid model has appeared in several areas of physics and mathematics. Indeed, the logarithmic repulsion in \eqref{eq:hamiltonian} does occur as interaction between vortices and dislocations in real systems such as superconductors \cite{Sandier12}, superfluids, rotating Bose-Einstein condensates \cite{Correggi08,Correggi12,Rougerie14} (we refer to \cite{Aftalion06} and \cite{Serfaty14} for detailed reviews). There is also a well-known analogy between the canonical measure \eqref{eq:jpdfC} of the 2D-OCP and the Laughlin trial wave function~\cite{Laughlin83} (in the symmetric gauge) in the theory of fractional quantum Hall effect~\cite{Can14}. In the large $N$ limit with fixed potential (without the factor $N$ in front of $V(\vec{r}/L)$), the 2D-OCP is equivalent to a class of growth models known as `Laplacian growth'~\cite{Weigmann06,Zabrodin09}.  

In the following $q$ and $L$ will be set to one for simplicity. For notational convenience, we also identify $\R^2\simeq\C$ and denote the positions of the particles in the plane by complex numbers $z_1,\dots,z_N$. With these conventions, \eqref{eq:jpdf}-\eqref{eq:hamiltonian} read 
\barr
&&\mathbb{P}_{\beta,N}\left(z_1,\dots,z_N\right)=\displaystyle\frac{1}{\mathcal{Z}_{N,\beta}}e^{-\beta H(z_1,\dots,z_N)}, \label{eq:jpdfC}\\
&&H\left(z_1,\dots,z_N\right)=\displaystyle-\frac{1}{2}\sum_{i\neq j}\log|z_i-z_j|+N\sum_kV\left(z_k\right). \label{eq:hamiltonianC}
\earr 

For $\beta=2$ and $4$, \eqref{eq:jpdfC} turns out to be the eigenvalues joint distribution for normal complex and normal self-dual matrix ensembles respectively~\cite{Zabrodin04,Khoruzhenko11,Bordenave12,Forrester10}. 
Remarkably, at inverse temperature $\beta=2$, \eqref{eq:jpdfC} is a determinantal point process. When the background fills uniformly the unit disk $D=\{|z|\leq1\}$, the plasma particles experience a quadratic confinement $V(z)=|z|^2/2$ and, for $\beta=2$,  the system is integrable (the partition function is known exactly for finite $N$). In this case, \eqref{eq:jpdfC} describes the eigenvalues distribution of the celebrated complex Ginibre ensemble \cite{Ginibre65}. The 2D Coulomb gas analogy has been extensively used in random matrix theory to investigate the large dimensional limit of random normal matrices  since the seminal works of Dyson \cite{Dyson62}.

The present paper deals with some macroscopic features of the 2D-OCP. We present explicit results for the disk configuration of the background. 
Our quantities of interest are the \emph{radial moments} $\Delta_N^{(p)}$ ($p>0$) of the 2D-OCP 
\be
\label{deltap}\Delta_N^{(p)}=
\begin{cases}
\displaystyle\frac{1}{N}\displaystyle\sum_{k}r_k^p&\text{if $0<p<\infty$}\ ,\\
\displaystyle\max_{k}r_k&\text{if $p=\infty$}\ ,
\end{cases}
\ee
where for notational convenience we have introduced polar coordinates $z=r e^{\mathrm{i} \theta}$ (hence $r_k=|z_k|$). The different normalizations for $p$ finite and $p=\infty$ have been chosen in order to have $\Delta_N^{(p)}=\mathcal{O}(1)$ for all $p$. 

The aim of this paper is to derive the asymptotic distribution of this class of radial observables 
in the limit of large number $N$ of particles at logarithmic scales\footnote{We use $a_N\approx b_N$ to denote $\log a_N\sim\log b_N$ for $N\to\infty$.} (i.e.\ including the large deviation tails). Our findings can be summarized as follows:
\begin{itemize}
\item[(i)] We show that the probability density function $\PP^{(\infty)}_N(x)=\langle \delta(\Delta_N^{(\infty)}-x)\rangle$ of the so-called \emph{edge density profile} $\Delta_N^{(\infty)}$ behaves for large $N$ as
\be
\PP_N^{(\infty)}(x)\approx
\begin{cases}
\exp\{-\beta N^2 \Psi^{(\infty)}_L(x)\},&\text{if $x\leq1$}\ ,\\
\exp\{-\beta N \Psi^{(\infty)}_{R}(x)\},&\text{if $x>1$}\ ,
\end{cases}\label{eq:largeN1}
\ee
where the left ($L$) and right ($R$) large deviation functions are computed explicitly (see~\eqref{eq:PsiL} and~\eqref{eq:PsiR} below). Note the different  exponential rates of suppression for $x\leq1$ ($\mathcal{O}(N^2)$) and $x>1$ ($\mathcal{O}(N)$). In the derivation of our results, it will be clear that: 
\begin{itemize}
\item[a)] The fluctuations for $x\leq1$ are driven by a collective behaviour of the charges; on the contrary, the fluctuations to the right $x>1$ are associated to a spontaneous symmetry breaking in the problem (the equilibrium configuration breaks the rotational invariance in the plane). This change of behaviour of the Coulomb gas is at the heart of the change of speed in the large deviations tails; 
\item[b)] Based on the analysis of the next-to-leading order corrections to~\eqref{eq:largeN1}, which we are also able to obtain (see~\eqref{eq:ntntl}), we show that a genuine $1/N$-expansion of the excess free energy for the 2D-OCP fails to exist if the fluid phase assumption is violated (i.e.\ when the plasma distribution becomes singular). Indeed, our computations show that, if the equilibrium distribution of the Coulomb gas has a singular component, the  large-$N$ expansion of  $\log\ZZ_{N,\beta}$ includes a \emph{non-universal} term of order $\mathcal{O}(N\log N)$; 
\item[c)] Studying the limit of $\psi_L^{(\infty)}(x)$ as $x\to 1^-$, we observe a rather unusual feature: the expected matching between the left rate function and the behaviour of the limiting distribution to the extreme left of $x=1$, which is customary in this type of problems, does not actually hold in this case.
\end{itemize}
\item[(ii)] For $p$ finite, we will study in detail the large $N$ behaviour of the moment generating function $\widehat{\PP}^{(p)}_N(s)=\langle e^{-\beta N^2 s\Delta_N^{(p)}}\rangle$. We compute explicitly the large $N$ limit up to the next-to-leading order term 
\be
\widehat{\PP}^{(p)}_N(s)=\exp\left\{-{\beta N^2}\left[E_{p}(s)+\frac{\beta-4}{4\beta N} S_{p}(s)+\dots\right]\right\}\ .\label{eq:largeN2}
\ee
The scaled logarithm of $\widehat{\PP}^{(p)}_N(s)$ is nothing but the excess free energy of the 2D-OCP. This free energy is given by the electrostatic energy term $E_p(s)$ at leading order, while the first $1/N$-correction is just a configurational entropy term $S_p(s)$. The explicit forms of $E_p(s)$ and $S_p(s)$ are given in~\eqref{eq:result_Ep} and~\eqref{eq:result_Sp}. For $p=1$ (\emph{mean radial displacement}) we recover recent results~\cite{Cunden15b}.  Amusingly, for $0<p<2$ the leading order term $E_p(s)$ has a weak non-analytic point at $s=0$ due to a change in topology of the plasma distribution. Furthermore, the \emph{order} of the phase transition (i.e. the order of the first non-continuous derivative of $E_p(s)$ at $s=0$), while being always at least $3$, depends on the parameter $p$ (the larger $p$ the weaker the transition and arbitrarily weak singularities at $s = 0$ are possible). These non-analyticities of the free energy at the ground state of the plasma unveil a non-regular behaviour of the high order cumulants of $\Delta_N^{(p)}$ that cannot be detected by a standard Gaussian approximation.

\end{itemize}
The paper is organized as follows. In section~\ref{sec:thermo} we discuss some aspects of the thermodynamic limit of the 2D-OCP and we introduce the main quantities of interest. Section~\ref{sec:RMT} contains a brief review on the role of the 2D-OCP in random matrix models where some finite-$N$ formulae are available for $\beta=2$. These results will be used to substantiate our findings. 
In section~\ref{sec:pinfinity} we present our results on the edge density profile $\Delta_{N}^{(\infty)}$.
In particular, in subsections~\ref{sub:1} and~\ref{sub:2} we demonstrate the split-off mechanism and we show the breakdown of the free energy $1/N$-expansion for a constrained 2D-OCP.  The large deviation functions of the radial moments $\Delta_{N}^{(p)}$ ($p$ finite) are considered in section~\ref{sec:results}. 
We conclude with some final remarks in section~\ref{sec:conclus}.

\section{Thermodynamics of the 2D-OCP}
\label{sec:thermo}
In order to describe the large $N$ limit of the 2D-OCP, it is useful to introduce the normalized density of the gas
\be
\mu=\frac{1}{N}\sum_{i=1}^N\delta_{z_i}\ .
\ee 
For large $N$, when the density approaches a macroscopic limit, the partition function $\ZZ_{N,\beta}$ of the 2D-OCP can be written as a functional integral over  densities 
\be
\ZZ_{N,\beta}=\int[\mathcal{D}\mu]e^{-\beta N^2 \A[\mu]}\label{eq:ZA}\ ,
\ee
where $\A$ is the so-called \emph{action} that can be determined as follows. Assuming a fluid phase for the 2D-OCP, one expects that the partition function admits a large-$N$ expansion~\cite{Weigmann03,Weigmann06}. Under this assumption, the leading order contributions to $\A$ are given by the electrostatic energy and the configurational entropy of the Coulomb gas. The energetic contribution comes from 
\be
\beta H(z_1,\dots,z_N)=\beta N^2\EE[\mu]+\frac{\beta N}{2}\int\de\mu(z)\log \ell(z)+\cdots\ ,\label{eq:mean-field_appr}
\ee
where $\EE[\rho]$ is an energy functional and $\ell(z)$ a short-distance cut-off.
The energy functional $\EE$ is the mean-field approximation of $H$,
\be
\EE[\mu]=-\frac{1}{2}\iint\limits_{z\neq z'}\de\mu(z)\de\mu(z')\log|z-z'|+\int\de\mu(z)V(z)\ .\label{eq:mean-field_ef}
\ee
In the continuous limit, this functional needs to be corrected by introducing a  cut-off that regularizes the particles self-interaction. This correction is provided by the second term in \eqref{eq:mean-field_appr}, where the short-distance cut-off $\ell(z)$ can be chosen as $\ell(z)^{-1}=\sqrt{N(\de\mu(z)/\de z)}$.
A second contribution to the action $\A$ arises from the Jacobian $J$ of the change of integration variables from the microscopic variables $z_k$'s to $\mu$: $\prod_{k}\de z_k=[\mathcal{D}\mu]N!e^{\log J[\mu]}$, where the factor $N!$ takes into account the symmetry under permutations of the charges.
A standard argument~\cite{Dyson62,Weigmann06} gives
\be
\log J[\mu]=-N\int\de\mu(z)\log\frac{\de\mu(z)}{\de z}\ .
\ee
Hence, by combining all the contributions we identify the action as 
\be
\A[\mu]=\EE[\mu]+\frac{\log N}{N}C_{\beta}+\frac{1}{N}\frac{\beta-4}{4\beta}\SSS[\mu]+\frac{1}{N}C^{\prime}_{\beta}+\cdots\ ,\label{eq:action_exp}
\ee
where $\EE[\mu]$ is the mean-field energy functional \eqref{eq:mean-field_ef},
\be
\SSS[\mu]=-\int\de\mu(z)\log(\de\mu(z)/\de z)\label{eq:entropy}
\ee
is the configurational entropy of the gas density, and $C_{\beta}$, $C^{\prime}_{\beta}$ cannot be determined by these macroscopic considerations. Note that the entropic term at order $1/N$ in the action vanishes at inverse temperature $\beta=4$.
The action representation \eqref{eq:ZA} of the partition function is amenable to a saddle-point approximation; for large $N$, the value of $\log\ZZ_{N,\beta}$ is given by the action $-\beta N^2\A[\mu_{\text{eq}}]$ evaluated at the equilibrium density $\mu_{\text{eq}}$, i.e. the minimizer of the mean-field energy functional $\EE$. The leading term of the action is then $\EE[\mu_{\text{eq}}]$. Lebl\'e and Serfaty~\cite[Eq.~(1.19)]{Serfaty15} have recently proved that, under regularity assumptions on the equilibrium measure $\mu_{\text{eq}}$, the coefficient of the next-to-leading term is universal and equal to  $C_{\beta}=-\beta/4$ (the universality of $C_{\beta}$ was already known for the 2D-OCP constrained on the line~\cite{Grava15,Ercolani03,Guionnet13}). In section~\ref{sec:pinfinity} we will see that this is not the case when $\mu_{\text{eq}}$ has singular components (non-fluid phase).  This explicit example shows that the assumption~\cite[Eq.~(2.7)]{Serfaty15} of $\mu_{\text{eq}}$ having a density (used to prove the universality of $C_{\beta}$) is pertinent and not only technical. 
The next correction for large $N$ includes the configurational entropy $\SSS[\mu_{\text{eq}}]$. Note, however, that if the  saddle-point density is \emph{not} in a fluid phase the configurational entropy~\eqref{eq:entropy} diverges, as $\de\mu_{\text{eq}}(z)/\de z$ is not finite on the set of singular points. The term $C^{\prime}_{\beta}$ has been recently investigated in~\cite{Serfaty15}; it has the physical interpretation of microscopic free energy (a weighted sum of specific entropy and `renormalized energy'). The microscopic free energy $C^{\prime}_{\beta}$ is universal~\cite[Eq.~(1.17)]{Serfaty15} (independent of $V$); however, there are no explicit results on its value for generic $\beta$. 

From now on, we focus on the harmonic potential case 
\be
V(z)=\frac{1}{2}|z|^2\ .\label{eq:harmonic}
\ee
As already mentioned in the introduction, this quadratic potential is generated by a uniform background in the unit disk $D=\{|z|\leq1\}$. The relevance of this choice in random matrix theory will be discussed later.  It is well-known that, in the limit of large number of particles, the 2D-OCP in the harmonic potential \eqref{eq:harmonic} concentrates in a bounded region of the plane. Not surprisingly, the limiting distribution of the plasma is the uniform distribution on the unit disk
\be
\mu_{\mathrm{disk}}=\frac{1}{\pi}\mathbf{1}_{D}\ ,\label{eq:disk}
\ee
which is exactly the configuration of the 2D-OCP that neutralizes the uniform background of opposite charge. This distribution on the disk is known as \emph{circular law} in random matrix theory~\cite{Bai97,Bordenave12,Girko84}. The electrostatic energy and the entropy of the gas are then easy to compute,
\be
\EE[\mu_{\mathrm{disk}}]=\frac{3}{8},\qquad \SSS[\mu_{\mathrm{disk}}]=\log\pi\ .\label{eq:valuesdisk}
\ee

For $\beta=2$, detailed results on the typical fluctuations of radially symmetric observables of this system have been obtained exploiting the aforementioned determinantal structure. In the next section, we summarize some of these results.

\section{The integrable case $\beta=2$ and random matrices}
\label{sec:RMT}
For $\beta=2$, the canonical distribution of the 2D-OCP~\eqref{eq:jpdfC} assumes a determinantal structure and provides a model whose partition function and correlation functions can be computed explicitly \cite{Chau98,Khoruzhenko11}. The probability measure in the complex plane (corresponding to $V(z)=|z|^2/2$),
\be
\mathbb{P}_{N,2}\left(z_1,\dots,z_N\right)=\frac{1}{\mathcal{Z}_{N,2}}\prod_{i<j}|z_i-z_j|^2\prod_k e^{-N|z_k|^2}\ ,\label{eq:ginibre}
\ee
is well-known in the theory of random matrices. Let $G=(G_{ij})_{i,j=1}^N$ be a $N\times N$ matrix whose entries are independent complex Gaussian with zero mean and unit variance. 
Ginibre \cite{Ginibre65} managed to show that  the joint distribution of the $N$ eigenvalues $z_1,\dots,z_N$ of $N^{-1/2}G$ is given by \eqref{eq:ginibre} with partition function $\mathcal{Z}_{N,2}=\left[1!2!\cdots N!\right]\pi^{N}N^{-\frac{N(N+1)}{2}}$.

Kostlan in~\cite{Kostlan92} managed to integrate out the angular variables $\theta_j$ of the eigenvalues $z_j=r_je^{\mathrm{i}\theta_j}$, 
and showed that, up to a random reshuffling, the eigenvalues moduli $r_j=|z_i|$ are distributed as a collection of independent $\chi$ random variables: $(r_1,\dots,r_N)\stackrel{d}{=}\sigma(\xi_1/\sqrt{N},\dots,\xi_N/\sqrt{N})$
where $\xi_1,\dots,\xi_N$ are independent positive random variables with density \footnote{In other words $\xi^2_k\stackrel{d}{=}\chi^2_{2k}/2$.}
\be
x\mapsto \frac{2}{\Gamma(k)}x^{2k-1}e^{-x^2},\qquad k=1,\dots,N\ , \label{eq:Kostlan}
\ee
and $\sigma$ is a random permutation uniformly distributed in $S_N$.

For $p=\infty$, using Kostlan's theorem \eqref{eq:Kostlan}, the following identity holds 
\begin{equation}
\Pr[\Delta_N^{(\infty)}\leq x]
=\prod_{k=1}^N\frac{\gamma(k,Nx^2)}{\Gamma(k)}\ ,\label{eq:gamma_inc}
\end{equation}
where $\gamma(k,y)=\int_0^y\de t\, t^{k-1} e^{-t}$ denotes the lower incomplete gamma function.
Differentiating~\eqref{eq:gamma_inc}, one gets a finite-$N$ formula for the probability density of $\Delta_N^{(\infty)}$
\be
\PP^{(\infty)}_N(x)=2Nxe^{-Nx^2}\Pr[\Delta_N^{(\infty)}\leq x]\sum_{k=1}^N\frac{(Nx^2)^{k-1}}{\gamma(k,N x^2)}\ .\label{eq:gamma_inc2}
\ee 
Using a quantitative version of the central limit theorem, Rider~\cite{Rider03} has proved the following remarkable limiting result for the typical fluctuations of $\Delta_N^{(\infty)}$. Let $\gamma_N=\log N-2\log\log N-\log2\pi$, $a_N=\sqrt{4N\gamma_N}$ and $b_N=1+\sqrt{\gamma_N/(4N)}$. 
Then, the rescaled variable $a_N(\Delta_N^{(\infty)}-b_N)$ converges in distribution as $N\to\infty$ to a standard Gumbel variable
\be
\lim_{N\to\infty}\Pr[a_N(\Delta_N^{(\infty)}-b_N)\leq x]=e^{-e^{-x}}\ .\label{eq:thmRider}
\ee
A similar result has been recently established \cite{Chafai14} for a general class of radially symmetric external potentials $V(z)=V(|z|)$. 
Rider's theorem~\eqref{eq:thmRider} describes the typical fluctuations of the `top eigenvalue' of complex Ginibre matrices. In this work, we address the question of atypical fluctuations. Few explicit results are available on large deviations for the 2D-OCP, and they are usually valid at the leading order in $N$, see~\cite{Allez14,Cunden15b,Jancovici93}.

Forrester~\cite{Forrester99} obtained a finite-$N$ formula for the moment generating function  of radially symmetric linear statistics. Specializing his result to $\Delta_N^{(p)}$ with $p$ finite one has
\be
\widehat{\PP}^{(p)}_N(s)=\prod_{\ell=1}^N\frac{1}{\Gamma(\ell)}\int_0^{\infty}\de t\,e^{-t-2sN(t/N)^{p/2}}t^{\ell-1}\ , \label{eq:finitelapl}
\ee
and from here it is possible to extract the average and variance of $\Delta_N^{(p)}$ at leading order in $N$
\be
\langle\Delta_N^{(p)}\rangle=\frac{2}{2+p},\qquad\mathrm{var}(\Delta_N^{(p)})=\frac{p}{4N^2}\ . \label{eq:cumulForr}
\ee
For a rigorous proof of central limit theorems at $\beta=2$ for radially symmetric observables (including $\Delta_N^{(p)}$ with average and variance as above) see \cite{Rider13}. Atypical fluctuations are not described, of course, within the Gaussian approximation.
Formulae~\eqref{eq:gamma_inc2},~\eqref{eq:finitelapl} and~\eqref{eq:cumulForr} will be used in the next sections to corroborate numerically our results, as well as to prove some of our claims for $\beta=2$.

\section{Large deviations of the edge density profile}
\label{sec:pinfinity}
The large $N$ distribution of the plasma $\mu_{\mathrm{disk}}$~\eqref{eq:disk} implies that the edge density profile $\Delta_N^{(\infty)}=\max_{k} r_k$ of the 2D-OCP converges to $x=1$ as $N\to\infty$. For $\beta=2$, it is known that the typical fluctuations of the edge, i.e. the `top eigenvalue' of Ginibre matrices, are described by a Gumbel distribution (see previous section). These typical fluctuations are of order $\mathcal{O}(1/\sqrt{N})$. Our goal is to find, for all $\beta>0$, the statistical law of the fluctuations of $\Delta_N^{(\infty)}$ of order $\mathcal{O}(1)$ (hence, atypical). This amounts to computing the following large $N$ limit
\be
-\lim\limits_{N\to\infty}\frac{1}{\beta N^2}\log \Pr[\Delta_N^{(\infty)}\leq x]\ .\label{eq:largeNlimit}
\ee
The above limit can be written as
\be
-\frac{1}{\beta N^2}\log \Pr[\Delta_N^{(\infty)}\leq x]=-\frac{1}{\beta N^2}(\log\ZZ_{N,\beta}(x)-\log\ZZ_{N,\beta})\ ,\label{eq:PZ}
\ee
where $\ZZ_{N,\beta}(x)$ is the partition function of the 2D-OCP with the constraint $r_k\leq x$ for $k=1,\dots,N$.
As discussed in section~\ref{sec:thermo},   $(-1/\beta N^2)\log\ZZ_{N,\beta}(x)$ is dominated for large $N$ by the saddle-point of the action in~\eqref{eq:ZA}. The saddle-point is given by the distribution of the plasma that minimizes the electrostatic energy functional 
\be
\EE[\mu]=-\frac{1}{2}\iint\limits_{z \neq z'}\de\mu(z)\de\mu(z')\log|z-z'|+\frac{1}{2}\int\de\mu(z)|z|^2\ , \label{eq:mean-field}
\ee
with the constraint that all charges are at distance at most $x$ from the origin. Therefore we have
\be
-\lim\limits_{N\to\infty}\frac{1}{\beta N^2}\log \Pr[\Delta_N^{(\infty)}\leq x]=-\frac{1}{\beta N^2}(\EE[\mu_x]-\EE[\mu_{\mathrm{disk}}])\ ,\label{eq:PZ2}
\ee
where $\mu_x$ minimizes $\EE[\mu]$ in~\eqref{eq:mean-field} with the constraints
\be
\mu\geq0\quad\text{and}\quad\int\limits_{|z|\leq x}\de\mu(z)=1\ .
\ee

Finding the minimizer of a logarithmic potential under constraints is usually a technically involved task. From electrostatic considerations, we know that the equilibrium distribution of the plasma concentrates on a compact set. In the interior of the support, the distribution has constant density $1/\pi$ and the excess of charges accumulates on the boundary of the support\footnote{This is true in general whenever the external potential is $V(z)=\frac{|z|^2}{2}+U(z)$ with $U(z)$ harmonic. Recall that, as for the more familiar Coulomb interaction in dimension $3$, at electrostatic equilibrium any excess of charge in a conductor is localized on the surface.}. 
Note that the support of the distribution is itself unknown. In our case, we can take advantage of the rotational symmetry of the problem and eventually we find
\be
\de\mu_x(r,\theta)=
\begin{cases}
\displaystyle\frac{\mathbf{1}_{r\leq x}}{\pi}\,r\de r\de\theta+\frac{(1-x^2)}{2\pi}\delta(x-r)\de r\de\theta,&\text{if $x<1$}\\
\displaystyle\mu_{\mathrm{disk}},&\text{if $x\geq1$}\ .
\end{cases}\label{eq:mux}
\ee
For $x\geq1$, the saddle-point density is independent of $x$  (the constraint on the 2D-OCP is ineffective). When $x<1$, the constraint forces the 2D-OCP in a smaller disk and a fraction of charge accumulates on the boundary of the disk. More precisely, in the interior of the disk of radius $x$ the density is constant and equal to $1/\pi$; the exceeding charge $(1-x^2)$ lies exactly on the boundary of the disk and, by symmetry, is uniformly distributed. Hence, for $x<1$ the equilibrium distribution has a singular component (the delta measure at distance $x=r$ in~\eqref{eq:mux}) and we say that the plasma is not in a fluid phase.
Evaluating $\EE[\mu_x]$ and using~\eqref{eq:valuesdisk} for $\EE[\mu_{\mathrm{disk}}]$ we  get
\be
-\lim\limits_{N\to\infty}\frac{1}{\beta N^2}\log \Pr[\Delta_N^{(\infty)}\leq x]=-\frac{\theta(1-x)}{8}\left(\log x^4+x^4-4x^2+3\right)\ ,
\label{eq:LD}
\ee
where $\theta(x)$ is the Heaviside theta function. Note that \eqref{eq:LD} is identically zero for $x\geq1$,
while for $x\to1^{-}$ it vanishes as $(x-1)^3$. Therefore the third-order derivative of the large deviation function~\eqref{eq:LD} is discontinuous at $x=1$. Similar non-analiticities in the large deviation function of  the edge density profile for 2D Coulomb gases in one dimension  have been recorded in previous works on Hermitian random matrices~\cite{Majumdar14}. Eq.~\eqref{eq:LD} is consistent with previous results on the large charge fluctuations of the one-component plasma (see~\cite[Eq. (2.6)]{Jancovici93} for $x\to1^{-}$, and \cite[Eq. (10)]{Allez14} that contains~\eqref{eq:LD} as a limiting case).

Inspired by the phenomenology of 2D Coulomb gases on the line~\cite{DePasquale10,Facchi13,Majumdar09,Majumdar14} we ask for the existence of the nontrivial limit
\be
-\lim\limits_{N\to\infty}\frac{1}{\beta N}\log\PP^{(\infty)}_N(x),\qquad \text{for }\,x>1\ ,\label{eq:ans}
\ee
where $\PP^{(\infty)}_N(x)=\langle \delta(\Delta_N^{(\infty)}-x)\rangle$ is the probability that the edge density takes value $x$. The idea is to compute this limit as energy cost of a configuration with one charge of the Coulomb gas at distance $x>1$. The intuition is that, for large $N$, the most probable configuration of the gas with at least one particle at distance $x>1$ is described by the (unperturbed) equilibrium measure $\mu_{\mathrm{disk}}$ (i.e., the circular law) with one single particle at distance $x$. Among all possible realizations of the constraint $\Delta_N^{(\infty)}=x$  (with $x>1$), this is the least energetic one \footnote{To our knowledge this idea has been proposed for Hermitian matrix models in~\cite[Theorem 2.2]{Johansson99} generalizing a previous result~\cite[Theorem 6.2]{Guionnet01} for the rate function of the largest eigenvalue of GOE matrices. Later, by means of Coulomb gas considerations and the split-off ansatz, the rate functions for Gaussian and Wishart matrices were computed explicitly in~\cite{Majumdar09}.}.  The presence of one particle outside the unit disk corresponds to a \emph{spontaneous symmetry breaking} in the system: the problem is symmetric under rotations $U(1)$ in the complex plane, however the plasma configuration with the split-off of a single particle is necessarily no longer symmetric.
The computation of~\eqref{eq:ans} therefore amounts to estimating the energy cost in pulling one particle away from its equilibrium position inside the unit disk $D$ and relocating it around a generic position $x e^{\mathrm{i}\theta}$ outside the disk ($x>1$). If the plasma distribution is $\mu_{\mathrm{disk}}$ then the electrostatic energy of a particle at $x e^{\mathrm{i}\theta}$ is
\begin{align}
-\int\de\mu_{\mathrm{disk}}(z)\log|z-xe^{\mathrm{i}\theta}|+\frac{1}{2}|xe^{\mathrm{i}\theta}|^2=
\begin{cases}
\displaystyle\frac{1}{2}&\text{if $x\leq1$},\\
\displaystyle-\log x+\frac{x^2}{2}&\text{if $x>1$}.
\end{cases}
\end{align}
Note that inside the unit disk ($x\leq1$) the electrostatic energy is constant (the gas is at equilibrium). 
The energy cost to transport a particle outside the unit disk at  distance $x>1$ is given by the energy difference, and hence
\begin{align}
 -\lim\limits_{N\to\infty}\frac{1}{\beta N}\log\PP^{(\infty)}_N(x)
&=-\log x+\frac{x^2}{2}-\frac{1}{2}\ .\label{eq:limit22}
\end{align}

\begin{figure}[t]
\centering
\includegraphics[width=.48\columnwidth]{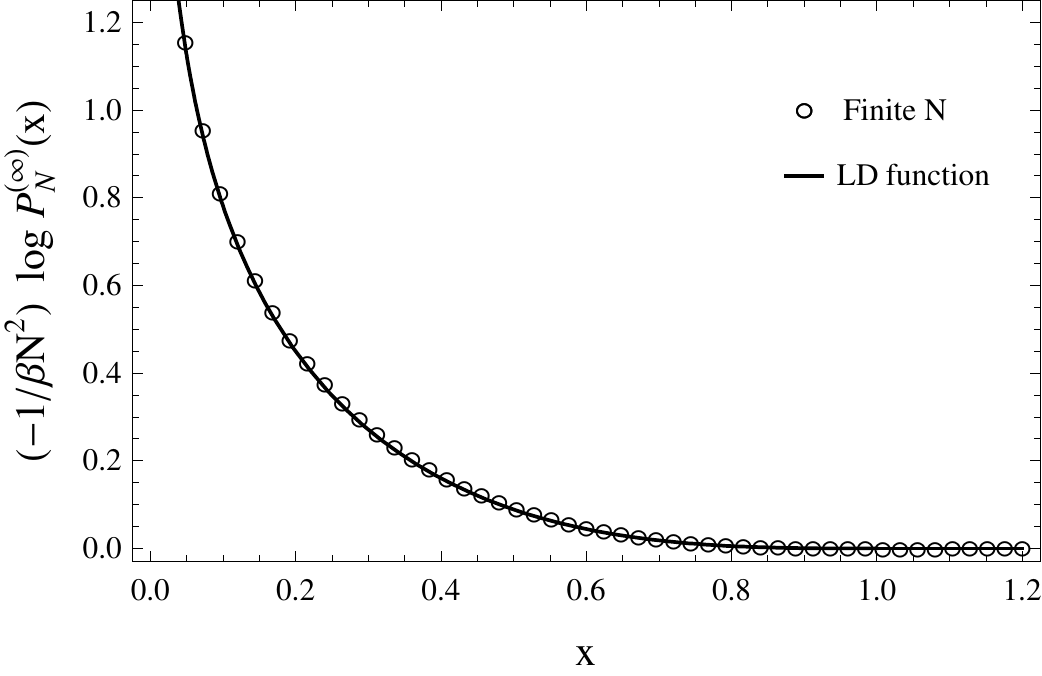}\quad
\includegraphics[width=.48\columnwidth]{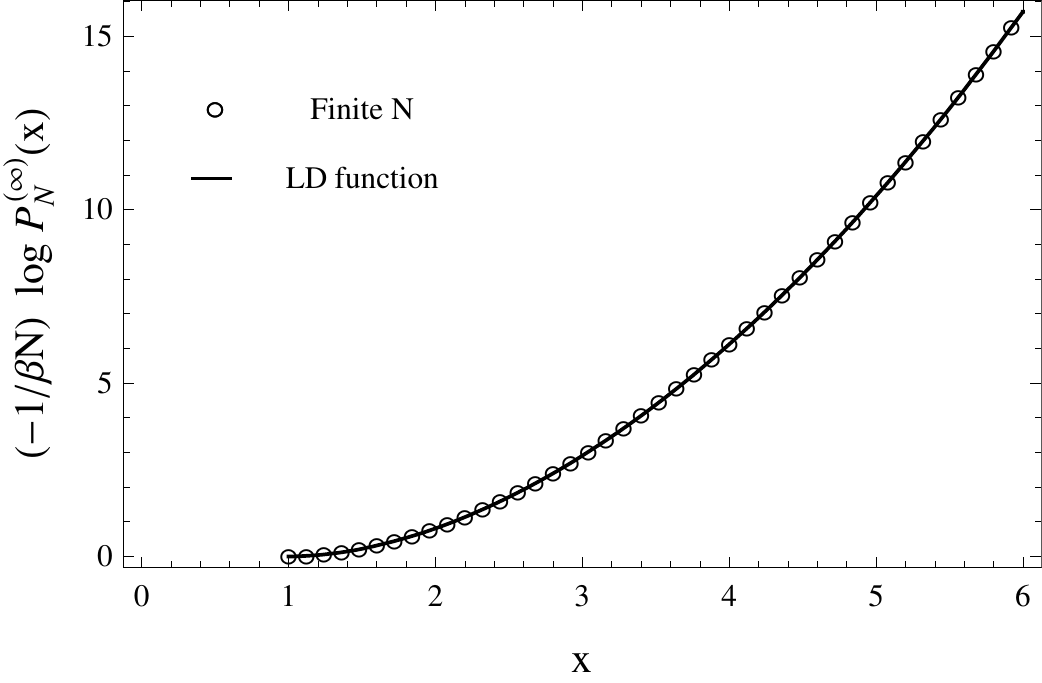}
\caption{\label{fig:LD} Comparison between the large deviation functions $\Psi^{(\infty)}_{L}(x)$ (left panel) in~\eqref{eq:PsiL} and $\Psi^{(\infty)}_{R}(x)$ (right panel) in~\eqref{eq:PsiR} with the finite-$N$ formula (dots) $\PP^{(\infty)}_N(x)$ for $\beta=2$ from~\eqref{eq:gamma_inc2} with $N=250$. }
\end{figure}

We can summarize the statistical fluctuation of the edge density $\Delta_{N}^{(\infty)}$ as in~\eqref{eq:largeN1} with left (L) and right (R) large deviation functions
\barr
\Psi^{(\infty)}_{L}(x)&&=-\frac{1}{8}\left(\log x^4+x^4-4x^2+3\right)\ ,\label{eq:PsiL}\\
\Psi^{(\infty)}_{R}(x)&&=-\frac{1}{2}(\log x^2-x^2+1) \label{eq:PsiR}\ .
\earr
For $x\leq1$ we have promoted the asymptotics~\eqref{eq:LD} of  the cumulative distribution $\Pr[\Delta_N^{(\infty)}\leq x]$ to a result for the probability density $\PP^{(\infty)}_N(x)$. There are  consistency conditions linking the two asymptotic formulae. Indeed, since $\Psi^{(\infty)}_{L}(x)$ is decreasing, using Laplace approximation we have $\Pr[\Delta_N^{(\infty)}\leq x]\approx\int_0^x \PP^{(\infty)}_N(t)\de t\approx \exp(-\beta N^2\inf_{t\leq x}\Psi^{(\infty)}_{L}(t))= \exp(-\beta N^2\Psi^{(\infty)}_{L}(x))$. Hence,  the cumulative distribution and the probability density have the same large-$N$ decay. We notice that the large deviation result~\eqref{eq:LD} does not exhibit a matching with the limiting result by Rider~\eqref{eq:thmRider} for typical fluctuations. This is due to the fact that the limiting (Gumbel) distribution~\eqref{eq:thmRider} decays too fast (super-exponentially) towards $-\infty$ to meet the left rate function at an exponential scale.
In Fig.~\ref{fig:LD} we show a comparison of these large deviation functions with the finite-$N$ formula~\eqref{eq:gamma_inc2} valid for $\beta=2$.

In the following subsections, we will corroborate our findings~\eqref{eq:PsiL}-\eqref{eq:PsiR} with an asymptotic expansion of the finite-$N$ formulae~\eqref{eq:gamma_inc}-\eqref{eq:gamma_inc2} valid for $\beta=2$. Moreover, the following analysis will i) demonstrate the split-off mechanism associated to atypically large values of $\Delta_N^{(\infty)}$, and ii) show that the excess free energy of a 2D-OCP constrained on a disk of radius smaller than $1$ does not have a genuine $1/N$-expansion: the presence of spurious logarithmic terms in the large-$N$ expansion is due to the constrained 2D-OCP being in a non-fluid phase.
\subsection{Proof of the split-off mechanism for $\beta=2$}\label{sub:1}
In the previous section we have computed the limit~\eqref{eq:ans} using large deviations ideas. A key step in the computation was finding the typical plasma distribution realizing the atypical event $\Delta_N^{(\infty)}=x$ with $x>1$. Guided by intuition, we have guessed that the typical way for the 2D-OCP to satisfy the constraint on the edge density profile is by a split-off mechanism. Evaluating the limit as energy cost for the splitting we found~\eqref{eq:limit22}. Here we obtain the \emph{same result} (i.e. $\Psi_R^{(\infty)}(x)$) using an asymptotic expansion of a finite-$N$ formula, thus proving the split-off phenomenon.

Let us consider the finite-$N$ formula~\eqref{eq:gamma_inc2} valid at $\beta=2$:
\be
\mathcal{P}^{(\infty)}_N(x)=2Nx e^{-Nx^2}\Pr[\Delta_N^{(\infty)}\leq x]\sum_{k=1}^N\frac{(Nx^2)^{k-1}}{\gamma(k,N x^2)}\ .
\ee
For $x>1$, we know that $\Pr[\Delta_N^{(\infty)}\leq x]\simeq1$ and, neglecting at most $o(N)$ terms, we write
\be
-\frac{1}{2N}\log\mathcal{P}^{(\infty)}_N(x)\sim\frac{x^2}{2}-\frac{1}{2N}\log \int_{1/N}^1\de t\, N\frac{(Nx^2)^{Nt-1}}{\gamma(Nt,N x^2)}\ ,\label{eq:approx1}
\ee
where the Euler-Maclaurin summation formula~\cite[Eq.~23.1.30]{Abramowitz70} has been used. It is easy to verify that $\log\gamma(Nt,N x^2)=tN\log N-N(t -t\log t)+o(N)$, for $t\leq1<x$, and hence we get
\be
-\frac{1}{2N}\log\mathcal{P}^{(\infty)}_N(x)\simeq\frac{x^2}{2}-\frac{1}{2N}\log\int_{1/N}^1\de t\,e^{-N[t\log t-t-t\log(x^2)]}\ .
\ee
Using a saddle-point approximation (the exponent achieves its minimum at the boundary $t=1$ of the integration domain) we eventually obtain the same result as in~\eqref{eq:limit22}
\be
\lim_{N\to\infty} -\frac{1}{2N}\log\mathcal{P}^{(\infty)}_N(x)=\Psi_R^{(\infty)}(x)\ .
\ee
This proves, at least for $\beta=2$, that the typical configuration of the 2D-OCP with edge density $\Delta_N^{(\infty)}>1$ is given by the unperturbed uniform distribution on the unit disk with one charge located at distance $x>1$ from the origin (hence one charge splits from $\mu_{\mathrm{disk}}$).

\subsection{Large-$N$ expansion for the constrained 2D-OCP}\label{sub:2}
We now consider the asymptotics of the free energy of the 2D-OCP confined in a disk of radius $x\leq1$. This amounts to considering the asymptotics of
\begin{eqnarray}
-\frac{1}{2N^2}\log\Pr[\Delta_N^{(\infty)}\leq x]
=-\frac{1}{2N^2}\sum_{k=1}^N\log\frac{\gamma(k,Nx^2)}{\Gamma(k)}\ .\label{eq:gamma_inc_app}
\end{eqnarray}
We have seen that the plasma under the constraint $r_k\leq x\leq1$ is not in a fluid phase (see~\eqref{eq:mux}). At equilibrium, the charges of the 2D-OCP accumulate on the boundary of the disk and the plasma distribution is not absolutely continuous. We want to show here that the excess free energy of the 2D-OCP with hard constraints does not admit a genuine $1/N$-expansion.

Using the Euler-Maclaurin summation formula
 we can write up to $o(N)$ error
\begin{align}
\log\Pr[\Delta_N^{(\infty)}\leq x]=\int_{1}^{N}\!\!\!\!\de t\log\left(\frac{\gamma(t,Nx^2)}{\Gamma(t)}\right)
+\frac{1}{2}\log\left(\frac{\gamma(N,Nx^2)}{\Gamma(N)}\right)+\ldots\ .\label{eq:EulerMac}
\end{align}
We have extracted the large-$N$ asymptotics of~\eqref{eq:EulerMac}.
Up to an error of order $o(N)$, we have the following asymptotics for large $N$  and $x\leq1$:
\begin{align}
\int_0^N\!\!\!\de t\log\gamma(t,N x^2)&\simeq\frac{N^2}{2}\log N+N\int_{1/N}^1\de\xi\,\log J_N(x,\xi)\ ,\label{eq:asymp1}\\
\int_{0}^{N}\!\!\!\de t\log\Gamma(t)&\simeq\frac{N^2}{2}\log N-\frac{3}{4}N^2-\frac{N}{2}\log N+\frac{1+\log(2\pi)}{2}N\ ,\label{eq:asymp2}\\
\log \gamma(N,Nx^2)&\simeq N\log N+N(\log x^2-x^2)\ ,\label{eq:asymp3}\\
\log \Gamma(N)&\simeq N\log N-N\ ,\label{eq:asymp4}
\end{align}
where $J_N(x,\xi)$ in~\eqref{eq:asymp1} is defined as
\be
J_N(x,\xi)=\int_{0}^{x^2}\frac{\de t}{t}e^{-N[t-\xi\log t]}\ .\label{eq:Jint}
\ee
The expression~\eqref{eq:Jint} lends itself to an evaluation in a saddle-point approximation. The largest contribution to the integral comes from the neighborhood of the point that minimizes the exponent $g_{\xi}(t)=t-\xi\log t$. This function achieves its global minimum at $t=\xi$, which may or may not be in the interior of the integration range $[0,x^2]$. 

Let us first consider the expansion in the case of $1/N< \xi<x^2$ (saddle-point in the interior of the integration range). Then, we can expand the exponential term around $t= \xi$ and obtain
\begin{align}
\ J_N(x,\xi)\simeq \frac{e^{-Ng_{\xi}(\xi)}}{\xi}\int_0^{x^2} \de t\,e^{-\frac{N}{2\xi}(t-\xi)^2} =e^{-N[\xi-\xi\log\xi]} \sqrt{\frac{2\pi}{N\xi}}\ .\label{eq:Jint1}
\end{align}
For $x^2<\xi\leq1$, the function $g_{\xi}(t)$ achieves its minimum at the boundary $t=x^2$ of the integration range and hence
\begin{align}
\ J_N(x,\xi)\simeq \frac{e^{-Ng_{\xi}(x^2)}}{x^2}\int_0^{x^2} \de t\,e^{-\frac{N(x^2-\xi)}{x^2}(t-x^2)} =\frac{e^{-N[x^2-\xi\log x^2]}}{N(x^2-\xi)}\ .\label{eq:Jint2}
\end{align}
Elementary integrations provide now the asymptotics of~\eqref{eq:asymp1} using the large-$N$ expansions~\eqref{eq:Jint1}-\eqref{eq:Jint2}:
\begin{align}
&\int_0^N\!\!\!\de t\log\gamma(t,N x^2)\simeq\frac{N^2}{2}\log N+N\int_{1/N}^{x^2}\!\!\!\!\de\xi\,\log J_N(x,\xi)+N\int_{x^2}^1\!\!\de\xi\, \log J_N(x,\xi)\nonumber\\
&=\frac{N^2}{2}\log N+\frac{N^2}{4} \left[x^4-4 x^2+4\log x\right]+\frac{N}{2}\log N \left[x^2-2\right]\nonumber\\
&+\frac{N}{2}\left[x^2(\log(2\pi)-1)-2x^2\log x+2+2(1-x^2)\log(1-x^2)\right]\ .\label{eq:asymp1fin}
\end{align}

\begin{figure}[t]
\centering
\includegraphics[width=.75\columnwidth]{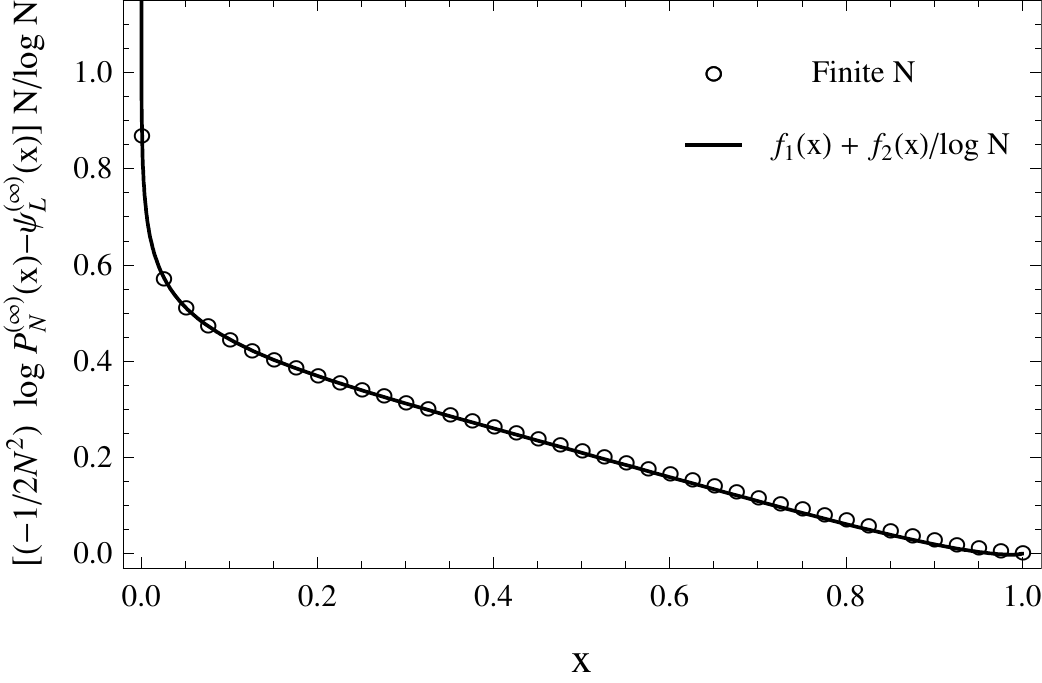}
\caption{\label{fig:figata} Comparison between $f_1(x)+f_2(x)/\log N$ of~\eqref{eq:f1}-\eqref{eq:f2} and the numerical evaluation of $(N/\log N)(-(1/\beta N^2)\log\Pr[\Delta_N^{(\infty)}\leq x]-\psi_L^{(\infty)}(x))$ at $\beta=2$ for $N=250$ and $x\leq1$.
}
\end{figure}

Collecting all terms in~\eqref{eq:asymp1fin} and~\eqref{eq:asymp2}-\eqref{eq:asymp4} we obtain the following large-$N$ expansion for $x\leq1$ (at $\beta=2$):
\barr
-\frac{1}{2 N^2}\log \Pr[\Delta_N^{(\infty)}\leq x]&=&-\frac{1}{2 N^2}(\log\ZZ_{N,2}(x)-\log\ZZ_{N,2})\nonumber\\
&=&\Psi^{(\infty)}_{L}(x)+\frac{\log N}{N}f_1(x)+\frac{1}{N}f_2(x)+\cdots
\ ,\label{eq:ntntl}
\earr
where $\Psi^{(\infty)}_{L}(x)$ is the left rate function~\eqref{eq:PsiL} and the correction terms are
\begin{align}
f_1(x)&=\frac{1-x^2}{4}\label{eq:f1}\ ,\\
f_2(x)&=\frac{1-x^2}{2}\left(\log(1-x^2)-\log{x}+\log\sqrt{2\pi}-1\right)\ . \label{eq:f2}
\end{align}
Note the nonzero coefficient of $\mathcal{O}(\log N/N)$.  
This explicit calculation shows that when the 2D Coulomb gas is not in a fluid phase (here for $x<1$), the coefficient $C_{\beta}$ in the large-$N$ asymptotics~\eqref{eq:action_exp} is not universal and therefore the excess free energy $\log (\ZZ_{N,\beta}(x)/\ZZ_{N,\beta})$ does not admit a genuine $1/N$-expansion. Note that here the hard constraint $r_k\leq x$ has more dramatic consequences on the free energy expansion of the 2D-OCP compared to the one-dimensional case (Hermitian random matrices)~\cite{Borot11,Chekhov06}.  For Hermitian matrix models, the large-$N$ eigenvalue density with a constraint on the largest eigenvalue is absolutely continuous~\cite{Borot11} and hence the term $\mathcal{O}(\log N/N)$ in the large deviation asymptotics is absent.
The expansion~\eqref{eq:ntntl} has been compared with the exact $\beta=2$ formula. See Fig.~\ref{fig:figata}. 

\section{Large deviations of radial moments}
\label{sec:results}

For finite $p$, it is more convenient to work in Laplace space and consider the moment generating function $\widehat{\PP}^{(p)}_N(s)=\langle e^{-\beta N^2 s\Delta_N^{(p)}}\rangle$. Taking the logarithm, we obtain the generating function of cumulants as excess free energy of the perturbed gas with respect to the unconstrained system
\be
-\frac{1}{\beta N^2}\log\widehat{\PP}^{(p)}_N(s)=-\frac{1}{\beta N^2}(\log\ZZ_{N,\beta}^{(p)}(s)-\log\ZZ_{N,\beta})\ ,\label{eq:PZ}
\ee
where $\ZZ_{N,\beta}^{(p)}(s)$ is the partition function of the 2D-OCP in a modified external potential
\be
V_s^{(p)}(|z|)=\frac{|z|^2}{2}+s|z|^p\ . \label{eq:mod_pot}
\ee
Using the large-$N$ representation \eqref{eq:ZA} of the partition function in terms of an action, one evaluates \eqref{eq:PZ}  as
\be
-\frac{1}{\beta N^2}\log\widehat{\PP}^{(p)}_N(s)=\A_s^{(p)}[\mu_s^{(p)}]-\A_0[\mu_{\mathrm{disk}}]\ ,\label{eq:excess2}
\ee
where  $\A_s^{(p)}$ is the action of the 2D-OCP in the external potential \eqref{eq:mod_pot} 
 and $\mu_s^{(p)}$ is the saddle-point density of $\A_s^{(p)}$. One readily sees that the saddle-point density $\mu_s^{(p)}$ is the equilibrium distribution of the plasma, i.e. $\mu_s^{(p)}=\arg\min\{\EE_s^{(p)}[\mu]\colon \mu\geq0\ , \int\de\mu(z)=1\}$ where
\be
\EE_s^{(p)}[\mu]=-\frac{1}{2}\iint\limits_{z\neq z'}\de\mu(z)\de\mu(z')\log|z-z'|+\int\de\mu(z)V_s^{(p)}(|z|)\ .
\ee

 The asymptotic behaviour of the cumulant generating function is obtained up to the next-to-leading order:
\be
-\frac{1}{\beta N^2}\log\widehat{\PP}^{(p)}_N(s)=E_{p}(s)+\frac{1}{N}\left(\frac{\beta-4}{4\beta}\right)S_{p}(s)+\cdots\ ,\label{eq:cumul_gen}
\ee
where
\begin{align}
E_p(s)&=\EE_s^{(p)}[\mu_s^{(p)}]-\EE_0[\mu_{\mathrm{disk}}]\ ,\label{eq:cumulGF}\\
S_p(s)&=\SSS[\mu_s^{(p)}]-\SSS[\mu_{\mathrm{disk}}]\label{eq:cumulGFntl}\ .
\end{align}
The leading term~\eqref{eq:cumulGF} is the rescaled cumulant generating function of $\Delta_N^{(p)}$, while $S_p(s)$ in~\eqref{eq:cumulGFntl} provides the $\mathcal{O}(1/N)$ correction to the cumulants. In this case, the $\mathcal{O}(N\log N)$ term in the expansion of $\log\ZZ_{N,\beta}^{(p)}(s)$ is universal (the potential $V_s^{(p)}(|z|)$ is sufficiently regular~\cite{Serfaty15}) and hence we have no logarithmic terms in the excess free energy~\eqref{eq:PZ}-\eqref{eq:cumul_gen}.

To summarize, the statistics of $\Delta_N^{(p)}$ is governed at leading order by the electrostatic energy excess $E_p(s)$ of the equilibrium distribution in the potential $V_s^{(p)}$ (independent of $\beta$) with the first correction in $1/N$ expressed in terms of a configurational entropy term $S_p(s)$.
Again, the technical challenge is the solution of a constrained optimization problem. In the next section, we solve the variational problem and we show that the statistical behaviour of $\Delta_N^{(p)}$ is quite rich and depends on the exponent $p>0$. 
\subsection{Solution of the variational problem}
Before tackling the optimization problem, notice that the moment generating function $\widehat{\PP}^{(p)}_N(s)$ is finite for $s\in\Omega_p$ with
\be
\Omega_p=
\begin{cases}
(-\infty,\infty),&\text{if $0<p<2$}\ ,\\
(-1/2,\infty),&\text{if $p=2$}\ ,\\
\left[0,\infty\right),&\text{if $p>2$}\ .
\end{cases}
\ee
These different ranges correspond to the stability values of $s$ for the 2D-OCP in the effective potential~\eqref{eq:mod_pot}. Observe that the equilibrium distribution $\mu^{(p)}_s$ will inherit the rotational symmetry of the energy functional $\EE_s^{(p)}$. In fact, the minimization problem can be solved explicitly using the arguments presented in~\cite{Cunden15b}.  A suitable modification of those steps provides the solution of the variational problem for $s\in\Omega_p$ as
\be
\de\mu^{(p)}_s(z)=\displaystyle\rho_s^{(p)}(r)\de r\frac{\de\theta}{2\pi}\mathbf{1}_{\Sigma_p(s)}\ , \label{eq:distreq}
\ee
with $\rho_s^{(p)}(r)$ given by
\be
\rho^{(p)}_s(r)=2r+sp^2r^{p-1}\ .\label{eq:rhos}
\ee
The equilibrium distribution $\mu^{(p)}_s$ of the plasma is supported on a radially symmetric domain
\be
\Sigma_p(s)=\left\{z\in\C\colon\, r_p(s)\leq |z| \leq R_p(s)\right\}\ ,
\ee
where $r_p(s)$ is the largest positive solution of $V_s'(r_p)=0$ and $R_p(s)$ is the unique positive solution of $V_s'(R_p)R_p=1$. These equations read explicitly
\begin{align}
& r_p=\max\{0,(-sp)^{\frac{1}{2-p}}\}\ ,\label{eq:rp}\\
& R_p^2+spR_p^{p}=1\ .\label{eq:Rp}
\end{align}
Note that as long as $s\geq0$ the effective potential \eqref{eq:mod_pot} is convex and therefore the plasma distribution is supported on a disk ($r_p(s)=0$). 
However, for $p<2$, when $s\in\Omega_p$ is negative the convexity of the potential is broken and the inner radius $r_p(s)$ becomes positive. Hence, for $s<0$ the support $\Sigma_p(s)$ of the plasma distribution is an annulus. This change of topology of the equilibrium distribution has important consequences on the statistical behaviour of the  radial moments.
The equilibrium measure $\mu^{(p)}_s(z)$ is the typical configuration of the 2D Coulomb gas with value of $\Delta_N^{(p)}$ given by
\be
x(s)=\int\de\mu^{(p)}_s(z)|z|^p=\frac{2}{2+p}\left(R_p^{p+2}-r_p^{p+2}\right)+\frac{sp}{2}\left(R_p^{2p}-r_p^{2p}\right)\ .
\ee
The electrostatic energy excess of $\mu^{(p)}_s$ can be conveniently evaluated using the identity~\cite[Eq.~(54)]{Cunden16}
\be
E_{p}(s)=\int_{0}^s\de s' x(s')\ ,
\ee
and eventually one finds for $s\in\Omega_p$
\begin{align}
E_p(s)&=\frac{1}{2}\int_{r_p}^{R_p}\de r\rho^{(p)}_s(r)V_s(r)+\frac{1}{2}(V_s (R_p)-\log R_p)-\frac{3}{8}\ ,\label{eq:result_Ep}\\
S_p(s)&=\int_{r_p}^{R_p}\de r\rho^{(p)}_s(r)\log(\rho^{(p)}_s(r)/r)-\log2\ .\label{eq:result_Sp}
\end{align}
Later we provide more explicit expressions for $E_p(s)$ and $S_p(s)$ for some values of $p$. 
We notice that, for $0<p<2$:
\be
E_p(s)=\text{(analytic part) }+ 
\begin{cases}
0&\text{if $s\geq0$}\\
c_p(-s)^{\frac{4}{2-p}}& \text{if $s<0$}\ ,
\end{cases}\label{eq:Ep_decomp}
\ee
where $c_p$ is a constant. Hence, $E_p(s)$ is not analytic at $s=0$ for $0<p<2$. Since $E_p(s)$ is the leading order in $N$ of the excess free energy~\eqref{eq:excess2}, this non-analyticities corresponds to \emph{phase transitions}. Note that the phase transitions occurs at the ground state of the 2D-OCP (at $s=0$) and correspond exactly to the disk-to-annulus change in topology of the equilibrium plasma distribution $\mu^{(p)}_s$ in~\eqref{eq:distreq}.  More precisely the $\ell$-th derivative of $E_p(s)$ is discontinuous when $2\left(1-\frac{2}{\ell-1}\right)<p\leq 2\left(1-\frac{2}{\ell}\right)$ with $\ell\geq3$. For instance, for $0<p\leq2/3$ we have a third-order phase transition, for $2/3<p\leq1$ a fourth-order phase transition, for $1<p\leq6/5$ the transition is of order five, and so on. For $p=2$ the function $E_p(s)$ is analytic. 
We stress that, although the underlying mechanism is the same (the change in topology), the order of the singularity in the free energies of the 2D-OCP depends on the particular statistics \footnote{The fact that the order of the phase transition depends on the observable, even if the macroscopic mechanism is the same, may appear surprising especially if compared with previous works for the 2D-OCP on the line (Hermitian random matrices, see e.g.~\cite{Majumdar14}). A general treatment of phase transitions for 2D Coulomb gases will appear in F. D. Cunden, P. Facchi, M. Ligab\'o and P. Vivo, in preparation.}, i.e. on the exponent $p$.

Having in mind these weak singularities, one can obtain the first cumulants $\kappa_{\ell}(\Delta^{(p)}_N)$ of $\Delta^{(p)}_N$ at leading order  in $N$ as derivatives of $E_p$ at $s=0$ (for $\ell<\frac{4}{2-p}$). For instance, one finds
\be
\kappa_1(\Delta_N^{(p)})=\frac{2}{2+p};\quad \kappa_2(\Delta_N^{(p)})=\frac{p}{2\beta N^2};\quad \kappa_3(\Delta_N^{(p)})=\frac{p^2}{2\beta^2N^4}\cdots\ .
\label{eq:cumulants}
\ee
For $\beta=2$, the obtained values of $\kappa_1(\Delta_N^{(p)})=\langle\Delta_N^{(p)}\rangle$ and $\kappa_2(\Delta_N^{(p)})=\mathrm{var}(\Delta_N^{(p)})$ concide with \eqref{eq:cumulForr}. The first correction to these results are obtained as derivatives at $s=0$ of $S_p(s)$. 

\subsection{Explicit formulae for $p=1$ and $p=2$}
For concreteness, we write here the form of $E_p$ and $S_p$ in terms of elementary functions for some special values of $p$. First, the expression of $E_p(s)$ can be simplified further as
\barr
E_p(s)&&=\frac{1}{8}\left(R_p^4-r_p^4\right)+\frac{4s+sp^2}{4(p+2)}\left(R_p^{p+2}-r_p^{p+2}\right)+\frac{s^2p}{4}\left(R_p^{2p}-r_p^{2p}\right)\nonumber\\
&&+\frac{1}{2}\left(\frac{R_p^2}{2}+sR_p^{p}-\log R_p-\frac{3}{4}\right)\ .
\label{eq:result_J}
\earr
For $p=1$ (the mean radial displacement) we obtain
\begin{align}
E_1(s)&=\frac{1}{2}\mathrm{arcsinh}\left(\frac{s}{2}\right)-\frac{s^2}{4}+\frac{s}{48}\left[\left(s^2+10\right)\sqrt{s^2+4}-|s|^3\right]\ ,
\label{eq:result_E1}\\
S_1(s)&=\mathrm{arcsinh}\left(\frac{s}{2}\right)-\frac{s^2+4}{8}\log(s^2+4)\nonumber\\
&+\frac{s}{4}\left[\sqrt{s^2+4}-s\log|s|-|s|\right]-\log2\ .
\label{eq:result_S1}
\end{align}
Eq. \eqref{eq:result_E1} was obtained in \cite{Cunden15b}.
Note that the \emph{fourth-order} derivative of $E_{1}(s)$ is discontinuous at $s=0$, in agreement with the previous discussion. 

\begin{figure}[t]
\centering
\includegraphics[width=.48\columnwidth]{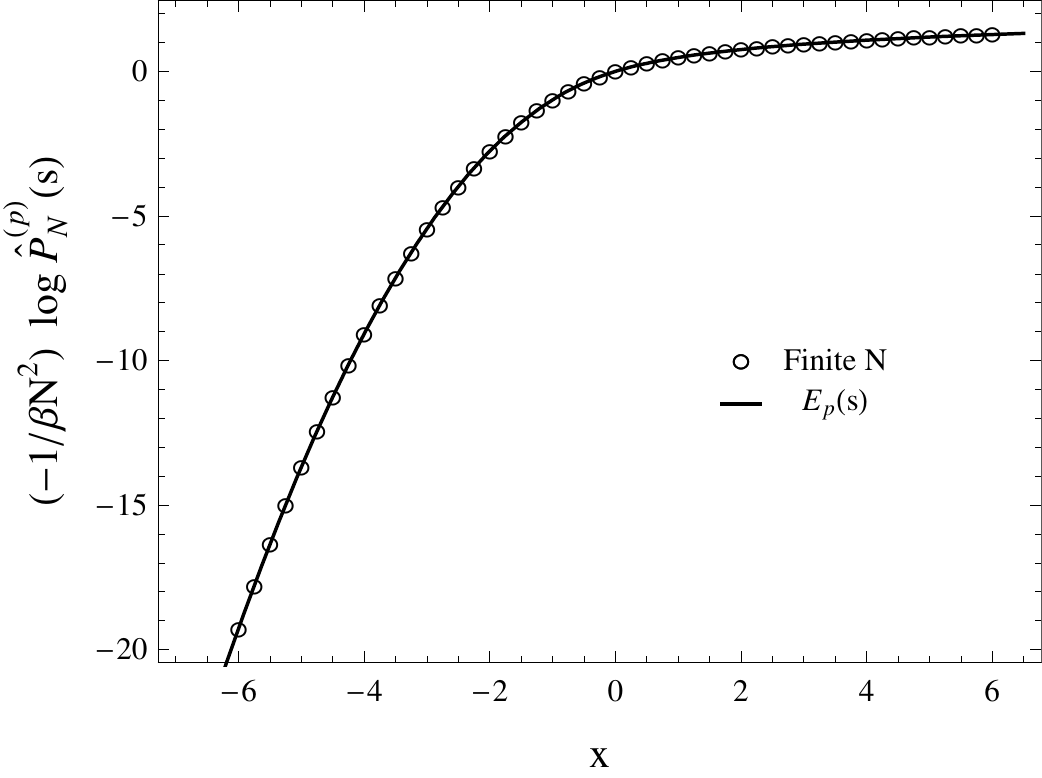}\quad
\includegraphics[width=.48\columnwidth]{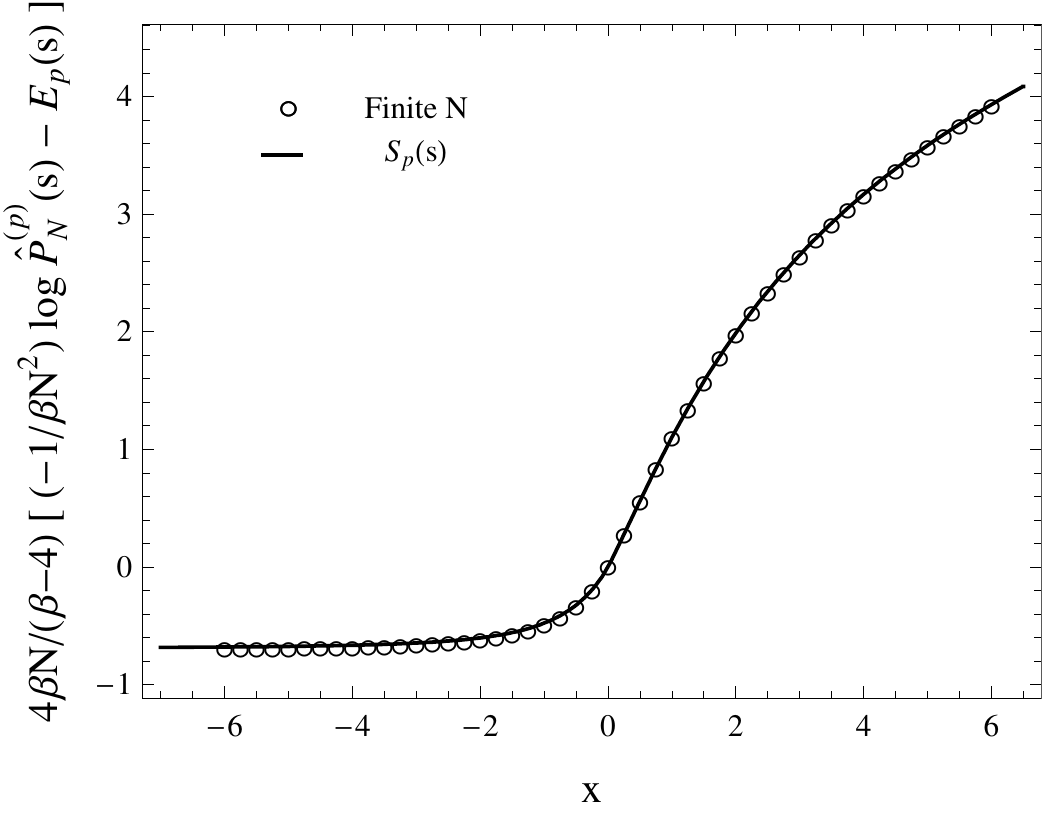}
\caption{\label{fig:Lapl} Large deviations of $\Delta_N^{(p)}$ for $p=1$. Comparison between the large deviation functions $E_p(s)$ (left panel) in~\eqref{eq:result_E1} and $S_p(s)$ (right panel) in~\eqref{eq:result_S1} with the finite-$N$ formula (dots) for $\beta=2$ from~\eqref{eq:finitelapl}. Here $N=50$.}
\end{figure}

\begin{figure}[t]
\centering
\includegraphics[width=.48\columnwidth]{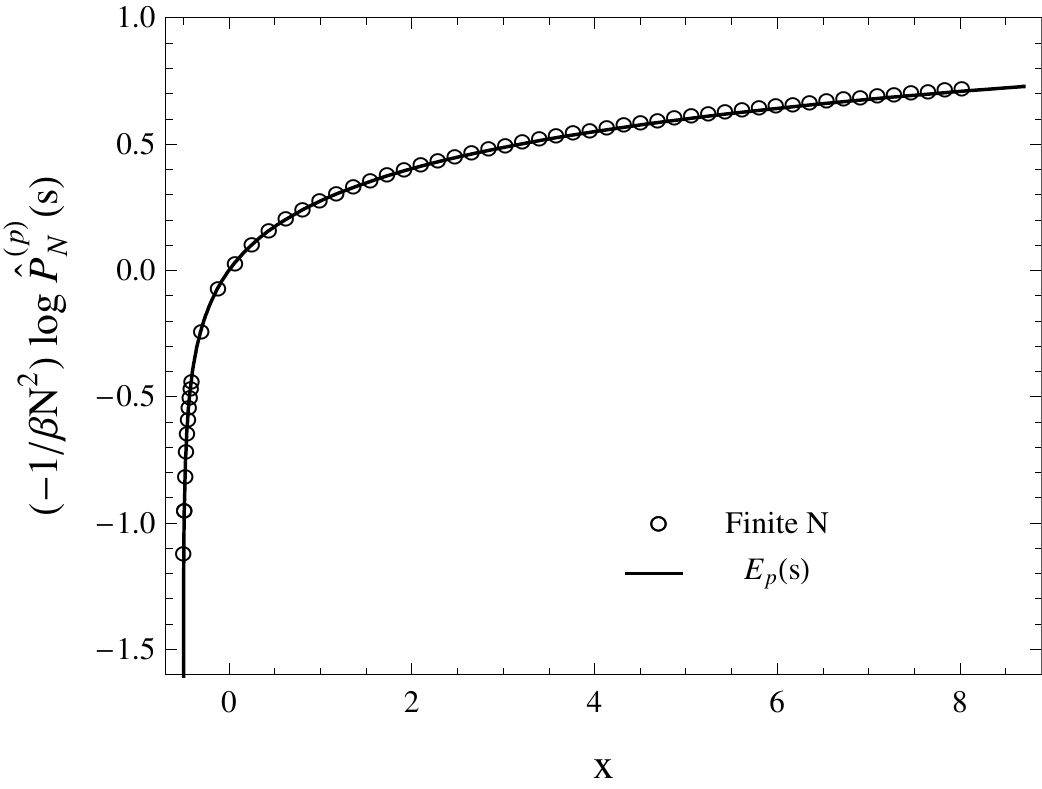}\quad
\includegraphics[width=.48\columnwidth]{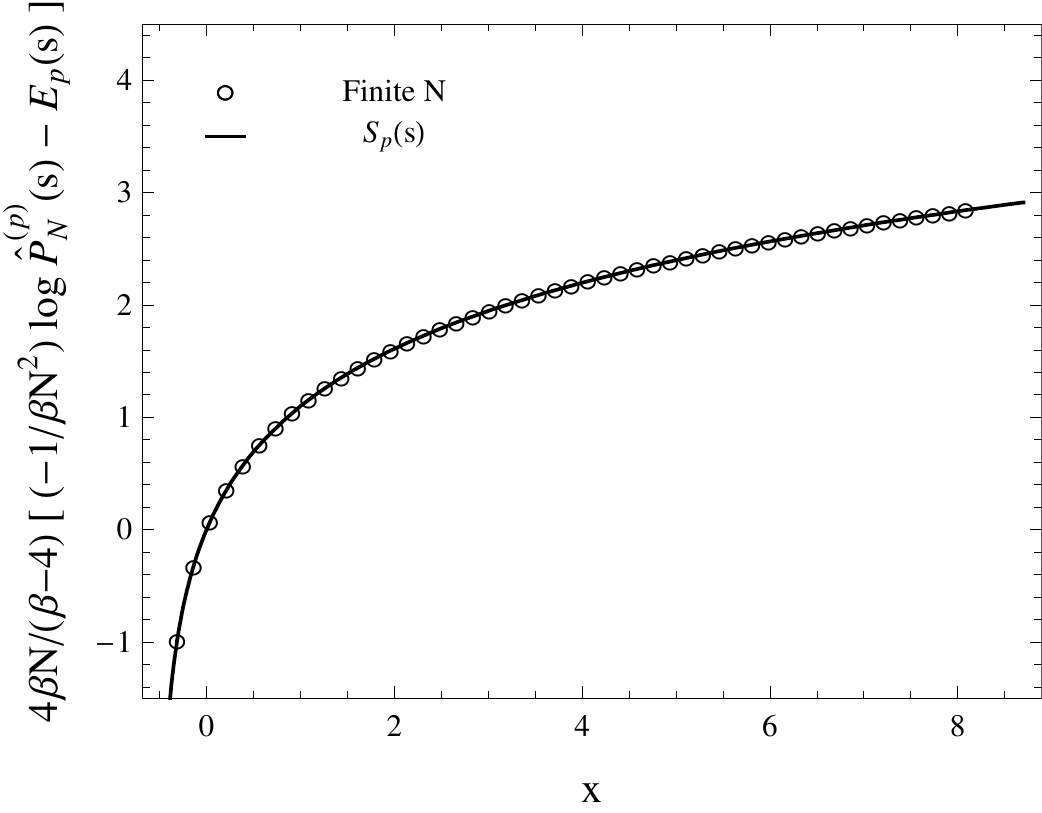}
\caption{\label{fig:Lapl2} Large deviations of $\Delta_N^{(p)}$ for $p=2$. Comparison between the large deviation functions $E_p(s)$ (left panel) in~\eqref{eq:result_E2} and $S_p(s)$ (right panel) in~\eqref{eq:result_S2} with the finite-$N$ formula (dots) for $\beta=2$ from~\eqref{eq:finitelapl}. Here $N=50$.}
\end{figure}

For the  moment of inertia $\Delta_N^{(2)}$ (i.e. $p=2$) we find the particularly simple expressions
\begin{align}
E_2(s)&=\frac{1}{4}\log(1+2s)\ ,\label{eq:result_E2}\\
S_2(s)&=\log(1+2s)\ .\label{eq:result_S2}
\end{align}
In this case $E_{2}(s)$ is real analytic, according to the fact that for $p=2$ there is no disk-to-annulus transition in the plasma distribution. In Figs.~\ref{fig:Lapl} and \ref{fig:Lapl2} we compare the asymptotic results~\eqref{eq:result_E1}-\eqref{eq:result_S2}  with the finite-$N$ expression~\eqref{eq:finitelapl} of $\widehat{\PP}^{(p)}_N(s)$ for $\beta=2$.

\section{Concluding remarks}
\label{sec:conclus}
In summary, we have obtained a number of results on radial observables $\Delta_N^{(p)}$ (as defined in Eq. \eqref{deltap}) for the two-dimensional one-component plasma in the limit of large number $N$ of particles and for generic values of the coupling constant. 

For finite $p$, the moment generating function of the $p$-th radial moment is obtained with logarithmic accuracy up to the first sub-leading term in the $1/N$-expansion. The leading order $E_p(s)$ of the excess free energy of the plasma is shown to exhibit a weak non-analytic point at $s=0$ for $0<p<2$: interestingly, the \emph{order} of this non-analyticity (i.e.  the order of the first non-continuous derivative), while being always larger or equal than $3$, is found to depend on $p$ and arbitrarily weak singularities at $s=0$ are possible. In the mean-field representation of the problem, these singularities are associated to disk-to-annulus transitions in the equilibrium distribution of the plasma.  Since $E_p(s)$ is the generating function of cumulants of $\Delta_N^{(p)}$ at leading order in $N$ we expect that the non-continuity of the derivatives of $E_p(s)$ at $s=0$ may be related to a `non-regular' behaviour of high cumulants as $N\to\infty$. For $p=1$ this was numerically observed in~\cite{Cunden15b}.

We have considered in detail the edge density profile $\Delta_N^{(\infty)}$ of the plasma in the  plane, i.e. the fluctuations of the farthest particle from the origin. We obtained the rate functions governing the probability of atypical fluctuations to the left and right of the expected position $x=1$. The difference in speeds ($\mathcal{O}(N^2)$ \emph{vs.} $\mathcal{O}(N)$) between the two large deviation principles is due to different arrangements of the fluid particles realizing an atypical configuration: large fluctuations to the left are realized cooperatively, while to the right by a splitting-off mechanism, which breaks the rotational symmetry of the fluid. This physically intuitive explanation has been also demonstrated for $\beta=2$ by the asymptotic expansion of the finite-$N$ formula~\eqref{eq:gamma_inc2}. To the best of our knowledge, this is one of the few existing proofs of the split-off phenomenon associated to the fluctuations of the maximum of a set of correlated random variables ($\{r_k\}$ in our problem).

 It is worth noticing that, at odds with what generally happens in this kind of problems, the large deviation result~\eqref{eq:LD} does \emph{not} exhibit a smooth matching with the limiting result by Rider (Eq.~\eqref{eq:thmRider}) for typical fluctuations to the left of the expected value, as the Gumbel distribution decays too fast (super-exponentially) towards $-\infty$ to meet the left rate function at an exponential scale. This problem gives then rise to another of the very few instances where the conventional matching between typical and atypical fluctuations is more delicate (see e.g.~\cite{Schehr13,PP15}).

A careful asymptotic analysis, extended to the first two sub-leading orders, also reveals that a genuine $1/N$-expansion of the excess free energy  requires a \emph{fluid-phase assumption}; in our problem the equilibrium density $\mu_x$ in~\eqref{eq:mux} is not absolutely continuous for $x<1$, and the large-$N$ expansion of $\log\Pr[\Delta_N^{(\infty)}\leq x]$ for $\beta=2$ contains a  nontrivial $\mathcal{O}(N\log N)$ term (see~\eqref{eq:ntntl}). This fluid-phase condition should be compared with the so-called \emph{one-cut assumption} for the 2D-OCP on the line~\cite{Guionnet13}. Note that, strictly speaking, the $1/N$-expansion also breaks down in Hermitian matrix models with a hard constraint of type $\lambda_{\mathrm{max}}\leq x$ (known as `hard wall'), $\lambda_{\mathrm{max}}$ being the largest eigenvalue. More precisely, one branch in the asymptotics of $\Pr[\lambda_{\mathrm{max}}\leq x]$ contains a nonzero term  $\mathcal{O}(\log N)$~\cite[Eq. (4.35)]{Borot11}. This term is only a `normalisation constant' independent of $x$, due to the presence of the hard constraint. Similar spurious $\mathcal{O}(\log N)$ contributions of geometric nature are also expected in two-dimensions.

It would be interesting to see if the results reported in this paper can be extended, at least qualitatively, to the  2D-OCP in more generic confining potentials \cite{Chau98,Khoruzhenko11,Zyczkowski02} (other than radially symmetric \cite{Akemann09,Girko84,Fyodorov03}) or for the plasma on different planar surfaces (e.g. on a cylinder~\cite{Can14}).

\begin{acknowledgements}
FDC and FM acknowledge  support  from EPSRC Grant\\ No.\ EP/L010305/1. FDC received furthermore partial support from the Italian National Group of Mathematical Physics (GNFM-INdAM). PV acknowledges the stimulating research environment provided by the EPSRC Centre for Doctoral Training in Cross-Disciplinary Approaches to Non-Equilibrium Systems (CANES, EP/L015854/1). FDC wishes to thank Anton Zabrodin and Gregory Schehr for stimulating discussions during the  `\'Ecole de Physique des Houches  2015 - Stochastic processes and Random matrices'. FDC is also grateful to Tamara Grava, Nicolas Rougerie and Maciej Nowak for clarifying discussions and for bringing relevant references to our attention.
\end{acknowledgements}


\end{document}